\documentclass[aps,prl,showpacs,showkeys,amsmath,amssymb,
twocolumn,
floatfix,
superscriptaddress
]{revtex4-1}
\usepackage{graphicx}
\usepackage{times}
\usepackage{verbatim}
\usepackage{color}

\newcommand{\nuebar}{$\overline{\nu}_{e}$}
\newcommand{\myol}[2][3]{{}\mkern#1mu\overline{\mkern-#1mu#2}} 
\newcommand{\TRSNOERROR}{0.090} 
\newcommand{\TRS}{\TRSNOERROR^{+0.008}_{-0.009}} 
\newcommand{\DMEE}{(2.59_{-0.20}^{+0.19}) \times 10^{-3}\ {\rm eV}^2} 
\newcommand{\DMNH}{(2.54_{-0.20}^{+0.19}) \times 10^{-3}\ {\rm eV}^2} 
\newcommand{\DMIH}{(2.64_{-0.20}^{+0.19}) \times 10^{-3}\ {\rm eV}^2} 
\newcommand{\TRO}{0.089\pm0.009} 
\newcommand{\TSO}{0.108\pm0.028} 
\newcommand{\DMEESO}{(2.55^{+0.21}_{-0.18}) \times 10^{-3}\ {\rm eV}^2}
\begin{document}

\title{Spectral measurement of electron antineutrino oscillation amplitude and frequency at Daya Bay}


\newcommand{\IHEP}{\affiliation{Institute~of~High~Energy~Physics, Beijing}}
\newcommand{\ECUST} {\affiliation{East China University of Science and Technology, Shanghai}}
\newcommand{\UW}{\affiliation{University~of~Wisconsin, Madison, Wisconsin, USA}}
\newcommand{\BNL}{\affiliation{Brookhaven~National~Laboratory, Upton, New York, USA}}
\newcommand{\NUU}{\affiliation{National~United~University, Miao-Li}}
\newcommand{\Dubna}{\affiliation{Joint~Institute~for~Nuclear~Research, Dubna, Moscow~Region}}
\newcommand{\CalTech}{\affiliation{California~Institute~of~Technology, Pasadena, California, USA}}
\newcommand{\CUHK}{\affiliation{Chinese~University~of~Hong~Kong, Hong~Kong}}
\newcommand{\NCTU}{\affiliation{Institute~of~Physics, National~Chiao-Tung~University, Hsinchu}}
\newcommand{\NJU}{\affiliation{Nanjing~University, Nanjing}}
\newcommand{\Tsinghua}{\affiliation{Department~of~Engineering~Physics, Tsinghua~University, Beijing}}
\newcommand{\SZU}{\affiliation{Shenzhen~Univeristy, Shenzhen}}
\newcommand{\NCEPU}{\affiliation{North~China~Electric~Power~University, Beijing}}
\newcommand{\Siena}{\affiliation{Siena~College, Loudonville, New York, USA}}
\newcommand{\IIT}{\affiliation{Department of Physics, Illinois~Institute~of~Technology, Chicago, Illinois, USA}}
\newcommand{\LBNL}{\affiliation{Lawrence~Berkeley~National~Laboratory, Berkeley, California, USA}}
\newcommand{\UCB}{\affiliation{Department of Physics, University~of~California, Berkeley, California, USA}}
\newcommand{\UIUC}{\affiliation{Department of Physics, University~of~Illinois~at~Urbana-Champaign, Urbana, Illinois, USA}}
\newcommand{\CDUT}{\affiliation{Chengdu~University~of~Technology, Chengdu}}
\newcommand{\SJTU}{\affiliation{Shanghai~Jiao~Tong~University, Shanghai}}
\newcommand{\BNU}{\affiliation{Beijing~Normal~University, Beijing}}
\newcommand{\WM}{\affiliation{College~of~William~and~Mary, Williamsburg, Virginia, USA}}
\newcommand{\Yale}{\affiliation{Department~of~Physics, Yale~University, New~Haven, Connecticut, USA}}
\newcommand{\VT}{\affiliation{Center for Neutrino Physics, Virginia~Tech, Blacksburg, Virginia, USA}}
\newcommand{\NTU}{\affiliation{Department of Physics, National~Taiwan~University, Taipei}}
\newcommand{\CIAE}{\affiliation{China~Institute~of~Atomic~Energy, Beijing}}
\newcommand{\UCLA}{\affiliation{University~of~California, Los~Angeles, California, USA}}
\newcommand{\SDU}{\affiliation{Shandong~University, Jinan}}
\newcommand{\NanKai}{\affiliation{School of Physics, Nankai~University, Tianjin}}
\newcommand{\UC}{\affiliation{Department of Physics, University~of~Cincinnati, Cincinnati, Ohio, USA}}
\newcommand{\DGUT}{\affiliation{Dongguan~University~of~Technology, Dongguan}}
\newcommand{\HKU}{\affiliation{Department of Physics, The~University~of~Hong~Kong, Pokfulam, Hong~Kong}}
\newcommand{\UH}{\affiliation{Department of Physics, University~of~Houston, Houston, Texas, USA}}
\newcommand{\Charles}{\affiliation{Charles~University, Faculty~of~Mathematics~and~Physics, Prague}}
\newcommand{\USTC}{\affiliation{University~of~Science~and~Technology~of~China, Hefei}}
\newcommand{\SYSU}{\affiliation{Sun~Yat-Sen~(Zhongshan)~University, Guangzhou}}
\newcommand{\Princeton}{\affiliation{Joseph Henry Laboratories, Princeton~University, Princeton, New~Jersey, USA}}
\newcommand{\RPI}{\affiliation{Department of Physics, Applied Physics, and Astronomy, Rensselaer~Polytechnic~Institute, Troy, New~York, USA}}
\newcommand{\CGNPG}{\affiliation{China~Guangdong~Nuclear~Power~Group, Shenzhen}}
\newcommand{\NDUT}{\affiliation{College of Electronic Science and Engineering, National University of Defense Technology, Changsha}} 
\newcommand{\ISU}{\affiliation{Iowa~State~University, Ames, Iowa, USA}}
\newcommand{\XJTU} {\affiliation{Xi'an Jiaotong University, Xi'an}}

\author{F.P.~An}\IHEP\ECUST
\author{A.B.~Balantekin}\UW
\author{H.R.~Band}\UW
\author{W.~Beriguete}\BNL
\author{M.~Bishai}\BNL
\author{S.~Blyth}\NTU 
\author{R.L.~Brown}\BNL
\author{I.~Butorov}\Dubna
\author{G.F.~Cao}\IHEP
\author{J.~Cao}\IHEP
% ...

\collaboration{The Daya Bay Collaboration}\noaffiliation
\date{\today}

\begin{abstract}
\noindent {A measurement of the energy dependence of 
    antineutrino disappearance at the Daya Bay Reactor Neutrino Experiment is
    reported.  Electron antineutrinos ($\overline{\nu}_{e}$) from six 
  $2.9$~GW$_{\rm th}$ reactors were detected with six detectors
  deployed in two near (effective baselines 512 m and 561 m) and
  one far (1579 m) underground experimental halls. Using 217 days of
  data, 41589 (203809 and 92912) antineutrino candidates were
  detected in the far hall (near halls).  An improved measurement of
  the oscillation amplitude  $\sin^{2}2\theta_{13} = 0.090^{+0.008}_{-0.009} $ and the first direct 
  measurement of the $\overline{\nu}_{e}$ mass-squared difference 
$|\Delta   m^{2}_{ee}|= (2.59_{-0.20}^{+0.19}) \times 10^{-3}\ {\rm eV}^2  $ is obtained using the observed 
$\overline{\nu}_{e}$ rates and energy spectra in a three-neutrino framework.
  This value of $|\Delta   m^{2}_{ee}|$ is consistent with $|\Delta     m^{2}_{\mu\mu}|$ measured by muon neutrino disappearance,
    supporting the three-flavor oscillation model.
}
\end{abstract}

\pacs{14.60.Pq, 29.40.Mc, 28.50.Hw, 13.15.+g}
\keywords{neutrino oscillation, neutrino mixing, reactor, Daya Bay}
\maketitle


Experimental measurements of neutrino oscillations have clearly established 
that neutrinos have mass and that the mass eigenstates mix~\cite{PDG}.
The Daya Bay experiment recently reported the discovery of the
disappearance of reactor antineutrinos over kilometer-long baselines, 
providing the most precise
measurement of the  mixing angle
$\theta_{13}$~\cite{DB, CPC}.
Other experiments have made consistent $\theta_{13}$ measurements~\cite{RENO,DC,Adamson:2013ue,Abe:2013xua}.
Precise knowledge of neutrino mixing and mass differences enables experimental
searches for CP violation, tests of the neutrino mass hierarchy and
precision tests of oscillation theory.  In
particular, the relatively large value of $\theta_{13}$ facilitates
a rich program of future neutrino oscillation
research~\cite{BIGT13,Li:2013zyd,Kettell:2013eos}.
It also allows the Daya Bay
experiment to report in this Letter an independent measurement of
the neutrino mass-splitting  via the distortion of the
reactor antineutrino energy spectrum.

In the framework of  three-flavor neutrino mixing in  vacuum, the
probability that an \nuebar\ produced with energy $E$ is detected as
an \nuebar\ at a distance $L$ is given by
%
\begin{eqnarray}\label{eq:psurv}
P_{\overline{\nu}_e\rightarrow\overline{\nu}_e} &=& 1 
-\cos^4\theta_{13}\sin^2 2\theta_{12}\sin^2\Delta_{21} \hfill \\
&-&\sin^2 2\theta_{13}(\cos^2\theta_{12}\sin^2\Delta_{31} + \sin^2\theta_{12}\sin^2{\Delta_{32}}) \nonumber ,
\end{eqnarray}
\noindent where $\Delta_{ji}\equiv 1.267 {\Delta}m^2_{ji}({\rm eV}^2)
\frac{L({\rm m})}{E({\rm MeV})}$, and ${\Delta}m^2_{ji}$ is the difference 
between the 
mass-squares of the mass eigenstates $\nu_j$ and $\nu_i$.  Since
$\Delta m_{21}^2\!\!\!\! \ll \!\!\!
\left|\Delta m_{31}^2\right|\!\!\!\approx\!\!\!\left|\Delta m_{32}^2\right|$~\cite{PDG},
the short-distance ($\sim$km) reactor \nuebar\ oscillation is due
primarily to the $\Delta_{3i}$ terms and naturally leads to the
definition of the effective mass-squared difference $\sin^2
\Delta_{ee} \equiv \cos^2\theta_{12}\sin^2
\Delta_{31}+\sin^2\theta_{12}\sin^2{\Delta_{32}}\,$~\footnote{Our $\Delta m_{ee}^2$ definition is consistent with 
H.~Minakata, H.~Nunokawa, S.~J.~Parke and R.~Zukanovich Funchal, Phys. Rev. D {\bf 74}, 053008 (2006).
}.

The Daya Bay experiment previously determined $\sin^22\theta_{13}$
using only the relative rates of \nuebar\ detected in three
antineutrino detectors (ADs) located near to and three ADs located far
from six nuclear reactor cores~\cite{DB, CPC}.  The effective mass
splitting $|\Delta m^{2}_{\mu\mu}|$ measured in $\nu_{\mu}$
disappearance~\cite{PhysRevLett.106.181801} provided a good
approximation of $|{\Delta}m^2_{ee}|$ in the rate-only measurement.
This Letter presents a combined analysis of the \nuebar\ rates and
energy spectra measured for the six detector data-taking period
from 24 December 2011 to 28 July 2012. This represents a 48\% increase
in statistics over the most recent result~\cite{CPC}.
The $\sin^22\theta_{13}$ uncertainty is reduced by inclusion of the
spectral information and the statistics of the complete six-AD data
period.  The spectral distortion due to the $\sin^2\Delta_{ee}$ term
provides  a strong confirmation that the observed \nuebar\ deficit is
consistent with neutrino oscillations and allows the first
direct measurement of $|{\Delta}m^2_{ee}|$.

A detailed description of the Daya Bay  experiment
can be found in~\cite{DBNIM,DBproposal}.
Each of the three experimental halls (EHs) contains functionally identical, three-zone
  ADs surrounded by a pool of ultra-pure water
segmented into two regions, the inner water shield (IWS) and outer water 
shield (OWS), which are instrumented with
photomultiplier tubes (PMTs).
In each AD, light created as a result of particle interactions
in the innermost zone, defined by an inner acrylic vessel (IAV) containing 
gadolinium-doped liquid scintillator (LS),
and the surrounding undoped  LS zone, is collected by 192 radially-positioned 20-cm PMTs
in the outermost mineral-oil region. The AD trigger threshold of 45 hit
PMTs or a summed charge of $\sim\!65$ photoelectrons in all PMTs corresponds
to about 0.4 MeV in the Gd-doped volume. The trigger inefficiency for
events above  0.7 MeV is negligible. Charge and
timing information for each PMT  are
available for energy calibration and reconstruction, as described
in Ref.~\cite{DBNIM}.  The detectors have
a light yield of $\sim\!165$ photoelectrons/MeV
and a reconstructed energy resolution of
$\sigma_E/E\approx8\%$ at 1 MeV.

Reactor antineutrinos are detected via the inverse $\beta$-decay (IBD) reaction,
$\overline{\nu}_{e} + p \to e^{+} + n$. The delayed
gamma rays (totalling $\sim\!8$~MeV) generated from the neutron capture on Gd with
a mean capture time of $\sim\!30~\mu{\rm s}$ enable  powerful
background suppression. The prompt light from the $e^{+}$ gives an
estimate of the incident \nuebar\ energy,
$E_{\overline{\nu}_e} = E_{\rm prompt} + \myol{E}_n + 0.78$~MeV,
where $E_{\rm prompt}$ is the prompt event energy including the positron kinetic energy and the
annihilation energy, and $\myol{E}_n$ 
is the average neutron recoil energy ($\sim\!10$~keV).

Interpretation of the observed prompt energy spectra requires characterization
of the detector response
 to $e^+$, $e^-$ and $\gamma$,
which maps the true energy ($E_{\rm true}$) to the
reconstructed energy ($E_{\rm rec}$).
$E_{\rm rec}$ is determined by scaling the measured 
total charge with a position-dependent correction~\cite{DBNIM,CPC}.
For a $\gamma$ or $e^-$, $E_{\rm true}$ is the kinetic energy; for a positron
$E_{\rm true}$ is the sum of the kinetic energy and the energy from annihilation.
The energy response is not linear due to scintillator
and electronics effects and is taken into account by two functions, $f_{\rm scint}$ and $f_{\rm elec}$, respectively.
The scintillator nonlinearity is particle- and energy-dependent, and is related to
intrinsic scintillator quenching and Cherenkov light emission.
The quenching effects are constrained by   standalone measurements with
a fast neutron beam as well as by 
neutron source data and radioactive $\alpha$-decays in the AD.
The Cherenkov contribution is also affected by absorption and reemission in the liquid scintillator.
The scintillator nonlinearity for electrons is described
by an empirical model
$f_{\rm scint}(E_{\rm true}) = E_{\rm vis}/E_{\rm true} = (p_0+p_3 \cdot E_{\rm true})/(1+p_1 \cdot e^{-p_2 \cdot E_{\rm true}})$,
where $E_{\rm vis}$ is the total visible light generated by the particle
and $p_i$ are the model parameters.
A GEANT4-based~\cite{Agostinelli2003250,1610988} 
Monte-Carlo simulation (MC) is used to relate the $e^-$
scintillator nonlinearity to the response for $\gamma$ and $e^+$.
The electronics nonlinearity, $f_{\rm elec}(E_{\rm vis})$, 
is introduced due to the interaction of the scintillation light time profile
and the charge collection of the front-end electronics.
Given the similar timing profiles for $e^\pm$ and $\gamma$s, 
it is modeled  as
an exponential function of $E_{\rm vis}$
as determined by
studying the time profile of charge in the data and MC.

The energy model, $f=f_{\rm scint}\times f_{\rm elec}$,
is  determined by a fit to
monoenergetic $\gamma$ lines from radioactive sources
and the continuous $\beta+\gamma$ spectrum extracted from  $^{12}$B data.
Sources were deployed at the center of all ADs   regularly
($^{68}$Ge, $^{60}$Co, $^{241}$Am-$^{13}$C)~\cite{DBNIM} 
and during a special calibration period in summer 2012
($^{137}$Cs, $^{54}$Mn, $^{40}$K, $^{241}$Am-$^{9}$Be, Pu-$^{13}$C) 
with AD1 and AD2 in near-hall EH1. 
In addition, gamma peaks in all ADs which could be identified with singles and
correlated spectra in data ($^{40}$K,  $^{208}$Tl, $n$ capture on H, C, and Fe) were
included. 
For source data with multiple gamma-line emissions, $f_{\rm scint}$ is computed 
for each gamma then summed up, whereas $f_{\rm elec}$ is computed based on
the total $E_{\rm vis}$. 
The $^{12}$B isotopes are produced cosmogenically at the rate of
about 900 (60) events/day/AD at the near (far) site.
The measured relative nonlinearity of $<\!0.3\%$ among 
6 ADs~\cite{CPC}  is negligible in the context of the energy model.

\begin{figure}[htb]
\includegraphics[width=\columnwidth]{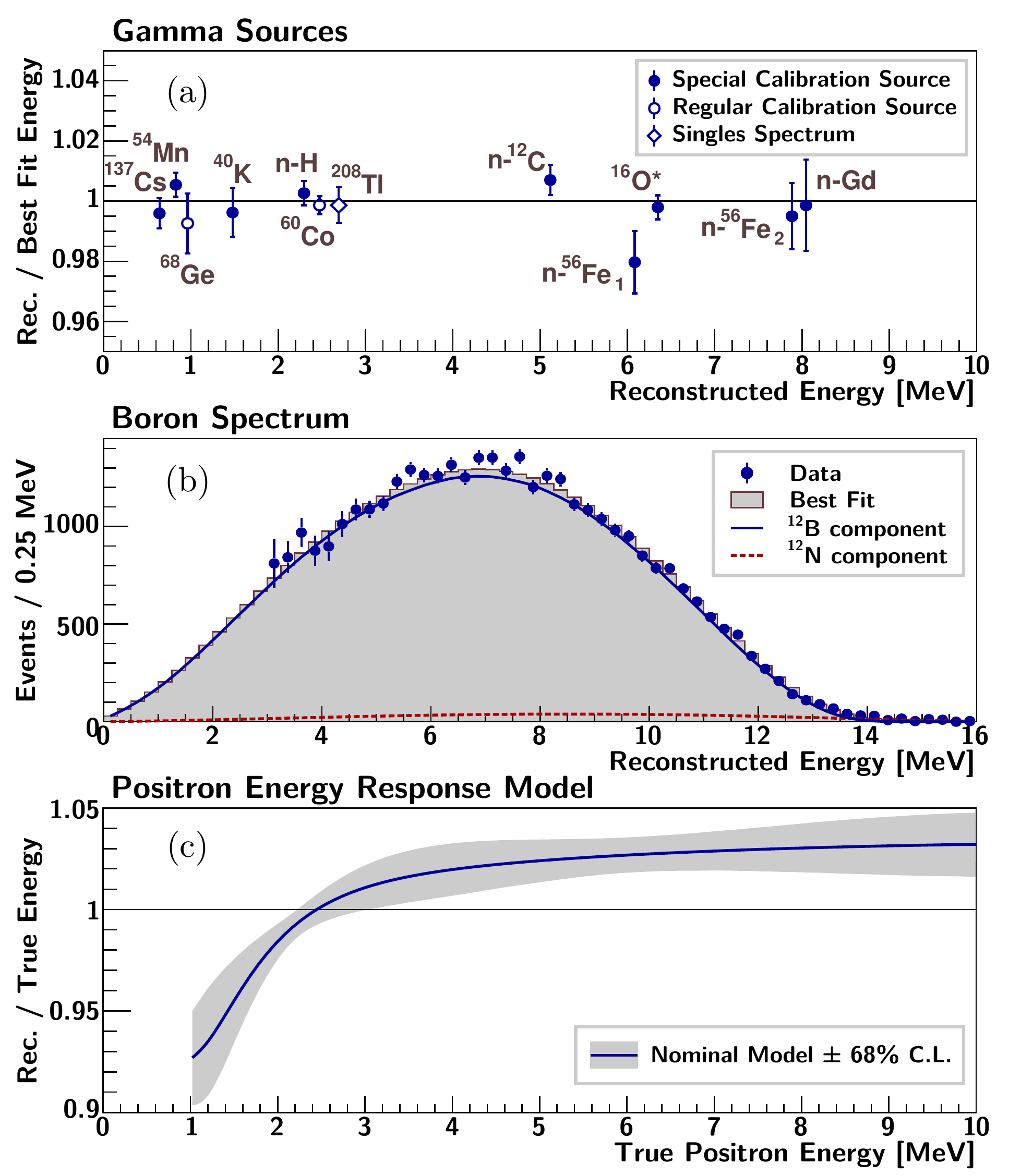}
\caption{
(a) Ratio of the reconstructed to best-fit energies of $\gamma$ lines
from calibration sources and singles spectra
as described
in the text.
The error bars represent the total uncertainty on each ratio.
The $\gamma$ from the second-excited state of ${}^{16}$O
in the Pu-$^{13}$C source is denoted  ${}^{16}{\rm O}^*$.
The n-${}^{56}{\rm Fe}_1$ and n-${}^{56}{\rm Fe}_2$ labels denote the
$\sim\!\!6$ MeV and $\sim\!\!7.6$ MeV $\gamma$s, respectively,  resulting
from the capture of neutrons from the AmC sources parked on top of the AD.
(b) Reconstructed energy spectrum (points) compared to the
sum (shaded area) of the $^{12}$B (solid line) and $^{12}$N (dashed line) components
of the best-fit energy response model.
The error bars represent the statistical uncertainties.
 (c) AD energy response model for positrons.
\label{fig:nonlin}}
\end{figure}

Figure~\ref{fig:nonlin} compares the best-fit energy model with the
 single-gamma, multi-gamma and 
continuous $^{12}$B   data used
to determine the model parameters.
As additional validation, the energy model prediction for the
continuous $\beta+\gamma$ spectra from $^{212}$Bi, $^{214}$Bi and $^{208}$Tl
decays was compared with the data and found to be consistent.

Alternative
energy response models, based on different methodologies, were constructed.
The second method builds the
scintillator nonlinearity based on Birks' formula~\cite{Birks} and Cherenkov radiation theory. The model is characterized by 
Birks' constant $k_B$ and the Cherenkov light contribution $k_c$. 
$f_{\rm elec}$ is determined from the residual 
nonlinearity of the same $\gamma$
and $\beta$-decay calibration data set.
The third method does not use $\gamma$ 
data but only uses  $\beta$-decay from $^{12}$B,
as well as the internal radioactive $\beta$-decays of $^{212}$Bi, $^{214}$Bi and $^{208}$Tl, to
construct the energy model.

All positron energy response models were  consistent
with each other to $\sim\!1.5\%$.
The uncertainty in the $e^+$ energy response, shown in Fig.~\ref{fig:nonlin}, is conservatively
estimated by combining the calibration and model uncertainties.
The energy response has a marginal effect on the measured oscillation parameters 
because it is essentially identical for all ADs.

The observed prompt energy spectrum is modified 
 because positrons from IBD interactions near the IAV can deposit
energy in the acrylic without generating scintillation light.
This significantly affects $\sim\!1\%$  of all IBD positrons causing an enhancement
near $E_{\rm rec}\approx 1$ MeV that is taken into account using MC.

The analysis used for previous Daya Bay results~\cite{DB,CPC} has been repeated
with the full six-AD data sample. The rate uncertainty of the background is 
slightly reduced compared to the previous analysis due to the increased 
statistics. The rate-only analysis yields 
$\sin^22\theta_{13} = 0.090\pm 0.010$ with $\chi^2/{\rm NDF} = 0.6/4$. 
The analysis has also been updated to 
include spectral information by applying the energy nonlinearity correction to 
the positron spectrum and measuring the spectral distribution of the five background
sources. 
The spectral uncertainties of the five backgrounds are 
included as uncorrelated among energy bins in the 
$\chi^2$ fit of the oscillation parameters, to allow 
all possible spectral models consistent with the data.
The combined rate and spectral analysis yields 
$\sin^22\theta_{13} = 0.092\pm 0.008$ and
$|\Delta m^2_{ee}| = (2.57^{+0.20}_{-0.22})\times 10^{-3} {\rm eV}^2$ 
with $\chi^2/{\rm NDF} = 166/171$ 
which are consistent with the results to be described in this Letter.

This Letter presents the results of an analysis that is largely 
independent of the  analysis described in \cite{DB,CPC}. 
 The two analyses differ in terms of event reconstruction,
energy calibration, IBD selection, background estimation
 and  construction
of the $\chi^2$ used for determination of the oscillation parameters.
The selected IBD candidates differ by  3.7\% (11\%) at the far (near) sites.
A ``blind analysis'' strategy was implemented by
concealing the reactor history and thermal power information for all cores
for the new data period.


 IBD candidates are selected with the criteria that follows.
First, events caused by PMT light emission are efficiently removed using the techniques of \cite{CPC}.
Candidates are then selected by requiring a prompt-like signal ($0.7-12$~MeV) in
coincidence with a delayed-like signal ($6-12$~MeV) separated 
by $1-200~\mu{\rm s}$.
Candidate pairs are vetoed if their delayed-like events occur
(i) within a ($-2~\mu{\rm s}$,~$600~\mu{\rm s}$) time-window
with respect to an IWS or OWS trigger with a PMT multiplicity  $>$12,
(ii) within a ($-2~\mu{\rm s}$,~$1400~\mu{\rm s}$) time-window
with respect to triggers in the same AD with a total light yield larger
 than $3000$~photoelectrons, or
(iii) within a ($-2~\mu{\rm s}$,~$0.4~{\rm s}$) time-window with respect to triggers in the
same AD with a total light yield higher than $3\times10^{5}$ photoelectrons.
This targeted muon veto allows for efficient removal of spurious
 triggers that follow a muon
as well as most muon-induced spallation products. Finally, a multiplicity cut is applied
to remove any ambiguities in the IBD pair selection. This cut requires no additional prompt-like
signals $400~\mu{\rm s}$ before the delayed event, and no  delayed-like signals
$200~\mu{\rm s}$ after the delayed event.
The muon veto efficiency ($\epsilon_\mu$) and multiplicity cut efficiency ($\epsilon_m$)
are calculated directly from data with negligible uncertainties for each AD.
The average values of $\epsilon_\mu \cdot \epsilon_m$ are summarized in Tab.~\ref{tab:ibd}.

A detailed treatment of the absolute and relative efficiencies, as well as 
their corresponding
uncertainties, has been reported in \cite{DBNIM, CPC}.
The uncertainties of
the absolute efficiencies are correlated among ADs and thus play a 
negligible role in the extraction of
the oscillation parameters. All differences among
ADs are treated
as uncorrelated uncertainties. In the rate-only analysis, the uncorrelated uncertainties are dominated
by the delayed-energy cut (0.12\%) and Gd capture fraction ($<$0.1\%).
In the spectral  analysis, additional uncorrelated uncertainty comes from
the relative energy scale difference between ADs. 
Based upon the relative response in all ADs to 
identified gamma and alpha peaks from numerous 
sources that span the IBD positron energy range, 
a 0.35\% uncertainty is assigned.


Five sources of background are identified. 
The accidental background,
defined as any pair of otherwise uncorrelated signals that happen to satisfy
the IBD selection criteria, is the largest background in the antineutrino sample.
The rate and energy spectra of this background can be accurately determined by
measuring the singles rates of prompt- and delayed-like signals and then
calculating the probability that the two randomly satisfy the selection criteria.
Alternative estimation methods yield consistent results.
The relative uncertainty of this background is 0.3\% 
and is dominated by the statistics
in  the rate of delayed-like signals.

The correlated $\beta-n$ decays from cosmogenic $^9$Li and $^{8}$He
can mimic IBD interactions.  The rate of correlated background from this source is estimated
by fitting the distribution of the time elapsed since the last muon with the
known $^9$Li and $^8$He decay lifetimes~\cite{Wen:2006hx}.
The 20\% systematic uncertainty takes into account the uncertainty
in $^9$Li and $^8$He production by muons with energy below the showering muon threshold. 
The rate is assumed to
be the same for ADs at the same site. The fraction of $^{9}$Li events in this
background is estimated to be $95\% \pm 5\%$ based on data and MC.
The spectra are calculated with a model that simulates the decay chain
of each isotope into their daughters based on
external data~\cite{Nyman1990189, Bjornstad:1981ad}.
The spectral uncertainty of this background
is estimated by assigning large variations to the energy response model,
particularly for the neutron and alpha daughter particles.

Neutrons from the $\sim\!\!0.7$~Hz Am-C calibration sources
inside the automated calibration units on top of the ADs
can occasionally mimic IBD events by inelastically scattering with
nuclei in the shielding material and then capturing on Fe/Cr/Mn/Ni.
This produces two $\gamma$ rays that both enter 
the scintillating region.
The MC is used to estimate the rate of this background.
The normalization is constrained by the measured rate of single delayed-like
candidates from this source.
A special Am-C source, approximately 80 times more potent than the calibration sources,
was temporarily deployed during summer 2012.
Results from this source are used to benchmark the MC 
and provide the estimate of the 45\% uncertainty
in the rate normalization.
The energy spectrum of this background 
is  modeled as an exponential, 
the parameters of which are constrained by these data.

Through elastic scattering with protons and the subsequent
thermalization and capture on gadolinium, energetic neutrons
produced by cosmic rays
can mimic IBD interactions. The energy of the proton-recoil signal
ranges from sub-MeV up to several hundred MeV.
If the prompt energy criterion is loosened to (0.7 - 50) MeV,
a flat spectrum is observed up to $50$~MeV,
which is extrapolated into the IBD energy region.
The flat spectrum assumption is corroborated
through the study of fast neutrons associated with muons identified
by the muon veto system  and by MC.
A $50\%$ systematic uncertainty in the rate is assigned.
The rate is assumed to be the same for ADs in the same experimental hall.

The $^{13}$C($\alpha$,n)$^{16}$O background is determined from a
simulation adjusted with the measured alpha-decay rates from
$^{238}{\rm U}$, $^{232}{\rm Th}$, $^{227}{\rm Ac}$
and $^{210}{\rm Po}$ decay chains. This background 
represents only about $0.01\%$ and $0.05\%$ of the total IBD sample in the near and far sites, respectively.

\begin{figure}[htb]
\includegraphics[width=\columnwidth]{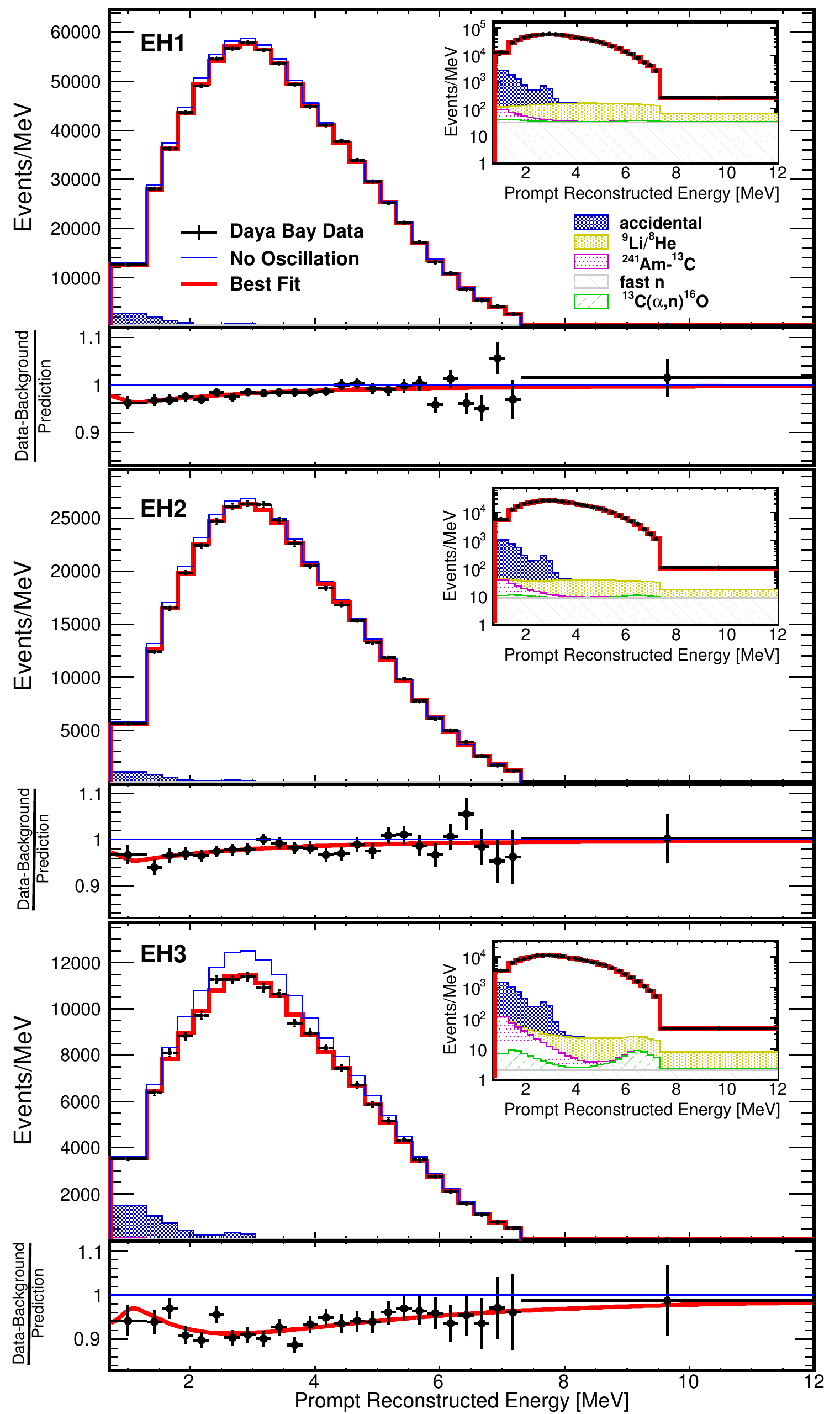}
\caption{The upper panel in each pair of panels shows the prompt positron
spectra (black points)  measured in the near (EH1 and EH2) and far (EH3) experimental halls with
the best-fit background contribution (shaded and colored regions). 
The thick red (thin blue) histograms represent the expected best-fit (no-oscillations) spectra.
The inset in each panel shows the same spectra with a logarithmic ordinate.
In the lower panel in each pair, the black points represent
the ratio of the background-subtracted data divided
by the predicted no-oscillation spectra. The error bars represent the statistical uncertainty only.
The red curve in each lower panel represents 
the ratio of the best-fit to no-oscillations spectra.
The change in slope of the red curve in the lowest energy bin is due to the effect of energy loss in the acrylic.
 \label{fig:farNearSpectralRatio}}
\end{figure}

The estimated IBD and background rates are summarized in Tab.~\ref{tab:ibd} and displayed
in Fig.~\ref{fig:farNearSpectralRatio}.
Backgrounds amount to about $5\%$ (2\%) of the IBD candidate sample in the far (near) sites.

\begin{table*}[!htb]
\begin{tabular}{|c|cc|c|ccc|}
\hline
                   & \multicolumn{2}{|c|}{EH1}&EH2&\multicolumn{3}{|c|}{EH3} \\
                  & AD1  & AD2  & AD3 & AD4 & AD5 & AD6 \\
\hline
IBD candidates &  101290 & 102519 & 92912 & 13964 & 13894 & 13731 \\
\hline
DAQ live time (days) & \multicolumn{2}{c|}{191.001}     &  189.645  & \multicolumn{3}{c|}{189.779}  \\
\hline
$\epsilon_{\mu}\cdot\epsilon_{m}$ &  0.7957 &  0.7927  & 0.8282  & 0.9577  & 0.9568 & 0.9566 \\
\hline
Accidentals (per day) &  9.54$\pm$0.03 &  9.36$\pm$0.03  & 7.44$\pm$0.02 & 2.96 $\pm$ 0.01 & 2.92 $\pm$ 0.01 & 2.87 $\pm$ 0.01 \\
\hline
Fast-neutron (per AD per day) &  \multicolumn{2}{c|}{0.92$\pm$0.46}   &  0.62$\pm$0.31   &  \multicolumn{3}{c|}{0.04$\pm$0.02} \\
\hline
$^9$Li/$^8$He (per AD per day) &  \multicolumn{2}{c|}{2.40$\pm$0.86}  & 1.20$\pm$0.63 & \multicolumn{3}{c|}{0.22$\pm$0.06}   \\
\hline
Am-C correlated (per AD per day) &  \multicolumn{6}{c|}{0.26$\pm$0.12}   \\
\hline
 $^{13}$C($\alpha$, n)$^{16}$O background (per day)&  0.08$\pm$0.04  &  0.07$\pm$0.04  & 0.05$\pm$0.03 & 0.04$\pm$0.02  & 0.04$\pm$0.02 & 0.04$\pm$0.02   \\
\hline
IBD rate (per day) &  653.30$\pm$2.31  & 664.15$\pm$2.33 & 581.97$\pm$2.07  & 73.31 $\pm$ 0.66 & 73.03 $\pm$ 0.66  & 72.20 $\pm$ 0.66 \\
\hline
\end{tabular}
\caption{Summary of signal and backgrounds. 
The background and IBD rates are corrected 
for the product of the muon veto and multiplicity 
cut efficiencies $\epsilon_{\mu}\cdot\epsilon_{m}$. \label{tab:ibd}  }
\end{table*}

The \nuebar\  spectrum from a reactor
with thermal power $W_{\rm th}(t)$ at 
energy $E$ and on a given day $t$ is 
\begin{align*}
\frac{d^2\!N(E,t)}{dEdt}=&\sum_{i}\left(\frac{W_{\rm th}(t)}{\sum_{j}f_{j}(t)e_{j}}\,
f_{i}(t)\, S_{i}(E)\, c^{\rm ne}_{i}(E,t)\right) \\ \nonumber
&+S_{\rm SNF}(E,t),
\end{align*}
with the fission fractions from each isotope $f_{i}(t)$, 
the thermal energy released per fission for each isotope $e_{i}$, 
the \nuebar\  yield per fission for each isotope $S_{i}(E)$, 
the correction to the \nuebar\  yield due to reactor 
non-equilibrium effects $c^{\rm ne}_{i}(E,t)$  and 
the spent nuclear fuel $S_{\rm SNF}(E,t)$. 
The nuclear reactor operators provide daily effective livetime-corrected 
thermal power  as well as periodic burn-up and simulation-based fission fraction
data  that are used to calculate daily fission fractions. The \nuebar\ 
flux at each detector is calculated by summing the contributions of
all reactors. 
The treatment of $W_{\rm th}$, $f_{i}$, $e_{i}$, $c^{\rm ne}_{i}$ and
$S_{\rm SNF}$ terms are described in \cite{DB,CPC}. 
The integrated, livetime-corrected, exposure for the EH3 ADs is 
$168.8\ {\rm kton}\!\cdot\!{\rm GW}_{\rm th}\!\cdot\!{\rm day}$ 
with mean fission fractions 
${}^{235}{\rm U}\!:\!{}^{238}{\rm U}\!:\!{}^{239}{\rm Pu}\!:\!{}^{241}{\rm Pu} = 0.573\!:\!0.076\!:\!0.301\!:\!0.050$.
Due to the relative measurement of near and far detectors, 
the measurement of oscillation parameters
is insensitive
to the choice of 
$S_i(E)$~\cite{Schrek1, Schrek2, Schrek3, Vogel1, HuberAnomaly, Mueller}.

\begin{figure}[htb]
\includegraphics[width=\columnwidth]{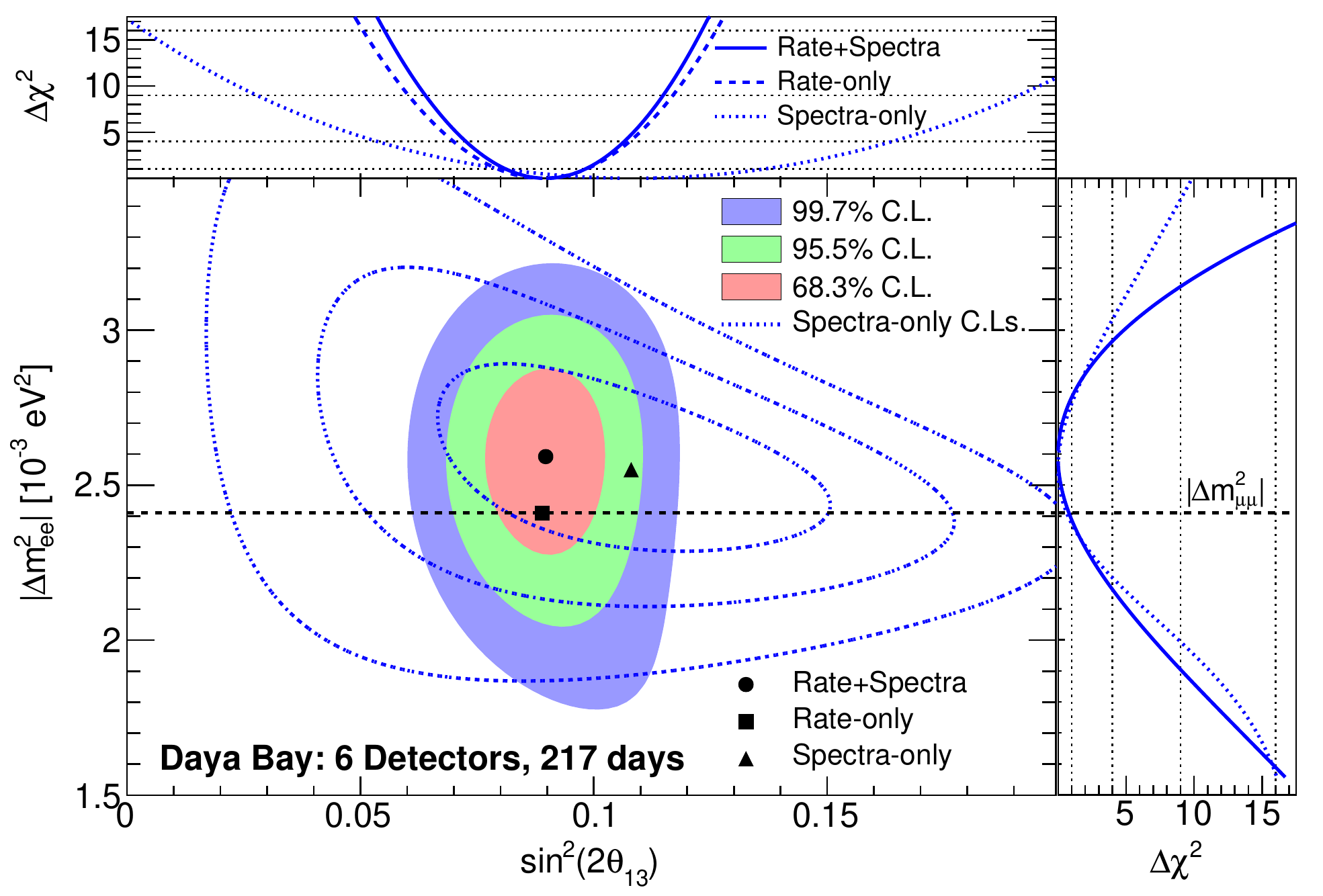}
\caption{
Allowed regions for the neutrino oscillation parameters
  $\sin^{2}2\theta_{13}$ and $|{\Delta}m^{2}_{ee}|$ at the 68.3, 95.5 and 99.7\%
confidence level, obtained from comparison of the rates
  and prompt energy spectra measured by the 3 near-site and 3 far-site
  antineutrino detectors (solid regions).  The best estimate of the
  oscillation parameters is given by the black dot.
The three dotted contours indicate the allowed 68.3, 95.5 and 99.7\%  C.L. regions for the spectra-only fit 
with the black triangle representing best estimate of the oscillation parameters.
 The adjoining panels show the dependence of $\Delta \chi^2$ on
$|{\Delta}m^{2}_{ee}|$ (right) and $\sin^22\theta_{13}$ (top).
The black square and dashed curve represent the rate-only result. 
The dotted curves represent the spectra-only $\Delta \chi^2$ distributions.
The dashed horizontal line represents the MINOS $|\Delta m^{2}_{\mu\mu}|$ measurement~\cite{Adamson:2013whj}.
 \label{fig:allowedOsc}}
\end{figure}

\begin{figure}[htb]
    \centering
\includegraphics[width=\columnwidth]{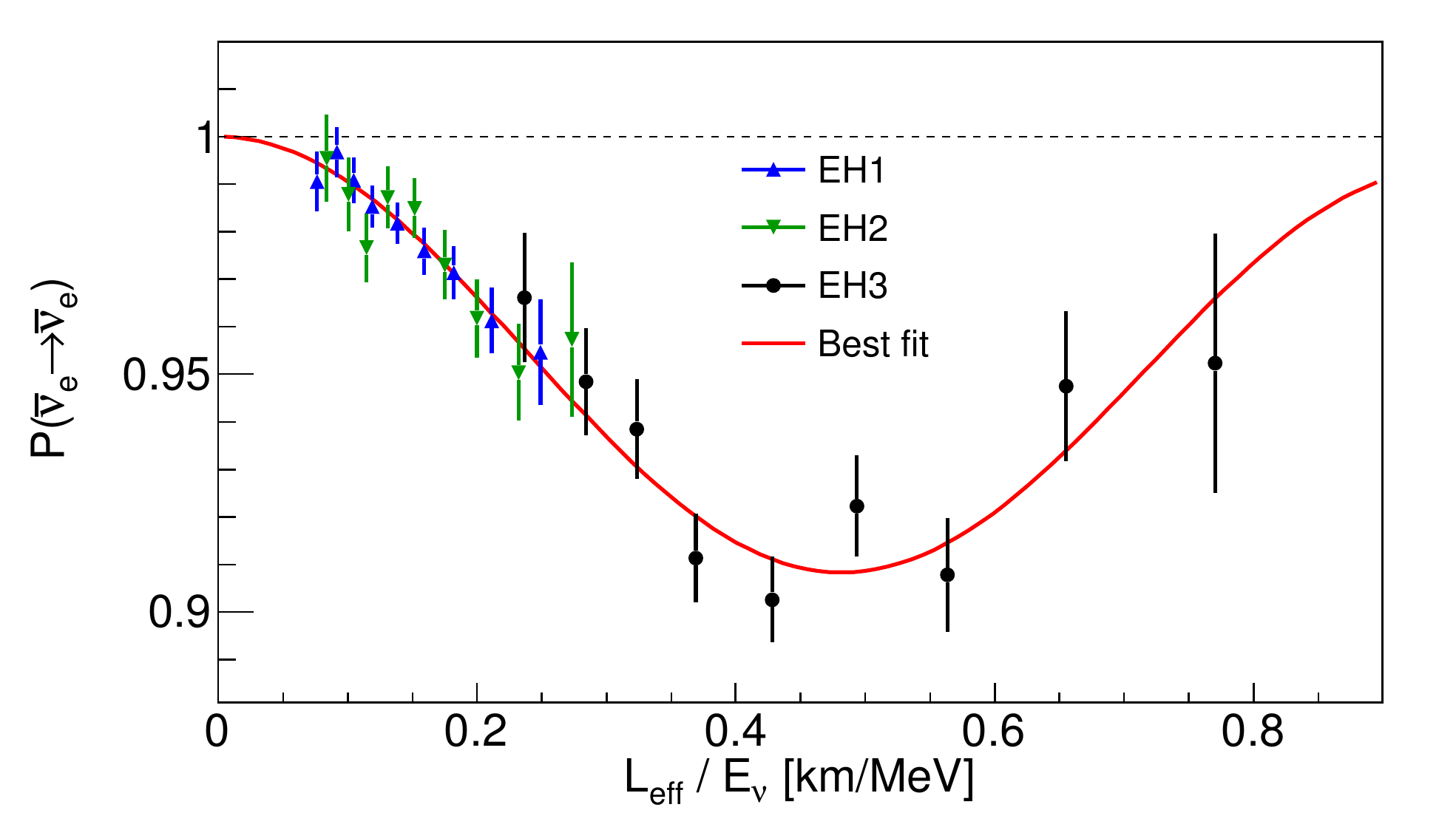}
    \caption{
Prompt positron energy spectra in the three experimental halls, 
re-expressed as the electron antineutrino survival probability versus 
propagation distance $L$ over antineutrino energy $E_{\nu}$.  
An effective 
detector-reactor distance $L_{\rm eff}$ is determined for each 
experimental hall equating 
the multi-core oscillated flux 
to an effective oscillated flux from a single baseline. 
The best estimate of the detector response is used to convert 
the background-subtracted positron energy spectrum 
into the antineutrino energy spectrum $E_{\nu}$.  
The horizontal location of each data point is given by the average of the counts 
in each bin ($\langle L_{\rm eff}/E_{\nu}\rangle$).  The vertical position is determined by the ratio 
of the counts in each bin relative to the counts expected assuming no oscillation, 
corrected for the reduction of 
analyzing power (energy dependent) due to 
multiple baselines and the binning in $L/E$.  
Error bars represent the statistical uncertainty only.  The oscillation survival 
probability using the best estimates of $\sin^22\theta_{13}$ and 
$|{\Delta}m_{ee}^{2}|$ is displayed for reference.  
}
    \label{fig:loe}
  \end{figure}

The oscillation parameters are extracted from a fit that takes into account the
antineutrino rate,  spectral  information and the \nuebar\ survival probability (Eqn.~\ref{eq:psurv}).
In order to properly account for the systematic effects and
correlations among energy bins, a $\chi^2$ is constructed using
nuisance parameters for detector response and background 
and a covariance matrix for reactor-related uncertainties. 
The absolute normalization of the \nuebar\ flux is a free parameter in the fit. 
The fit uses
$\sin^22\theta_{12}=0.857 \pm 0.024$ and
$\Delta m^{2}_{21}=(7.50 \pm 0.20)\times10^{-5} {\rm eV}^2$~\cite{PDG}.
The best-fit values are $\sin^{2}2\theta_{13} = \TRS $
and $|\Delta m^{2}_{ee}|=\DMEE $ 
with  $\chi^2/{\rm NDF} = 163/153$ 
(68.3\% confidence level (C.L.) intervals)~\footnote{See Supplemental Material at [URL will be inserted by publisher] for a table of  $\chi^2-\chi^2_{\rm min}$ as a function of $(\sin^22\theta_{13}$, $|\Delta m_{ee}^2|)$.}.
The prompt energy spectra  observed in each of the experimental halls  are 
compared to the spectra expected for no oscillation and with the best-fit oscillation parameters
in Fig.~\ref{fig:farNearSpectralRatio}.
The $68.3\%$, $95.5\%$, and $99.7\%$ C.L. 
allowed regions in the $|{\Delta}m^{2}_{ee}|$ vs. sin$^{2}2\theta_{13}$ plane
are shown in Fig.~\ref{fig:allowedOsc}.
Under the assumption of the normal (inverted) neutrino mass hierarchy~\cite{PDG},
this result is equivalent to 
$|\Delta m^{2}_{32}|=\DMNH $ ($|\Delta m^{2}_{32}|=\DMIH $).
The result is consistent with
$|{\Delta}m^{2}_{\mu\mu}| = ({2.41}^{+0.09}_{-0.10})\times 10^{-3}\ {\rm eV}^2$
as measured via $\nu_\mu$ and $\overline\nu_\mu$ disappearance~\cite{Adamson:2013whj}
noting  the small ${\cal O}(0.04\times10^{-3}\ {\rm eV}^2)$
effects due to other neutrino
oscillation parameters.
Figure~\ref{fig:loe} compares the IBD data from all experimental halls with  the
\nuebar\ survival probability (Eqn.~\ref{eq:psurv}) using the best-fit values.
Almost one full oscillation cycle is visible, 
demonstrating both the amplitude and frequency of short-baseline 
reactor $\overline{\nu}_e$ oscillation.

The total uncertainty on both oscillation parameters is 
dominated by statistics.
The most significant contributions to the $\sin^22\theta_{13}$
systematic uncertainty are the  reactor, relative-detector-efficiency
and energy-scale 
components~\footnote{Define the contribution of the $i^{\rm th}$ component 
to the total uncertainty ($\sigma_{\rm tot}$) as $\sigma_i^2/\sigma_{\rm tot}^2$. 
The contributions to the $\sin^22\theta_{13}$ ($|\Delta m_{ee}^2|$) uncertainty 
are then 0.73 (0.65), 0.18 (0.02), 0.13 (0.21), 0.11 (0.01), and 0.04 (0.06) 
for the statistical, reactor, relative-energy and efficiency, absolute-energy, 
and background components, respectively. 
Note that $\sum_i \sigma_i^2/\sigma_{\rm tot}^2\neq 1$ due to correlations.}.
The $|\Delta m^{2}_{ee}|$ systematic uncertainty is dominated by the 
relative energy scale and efficiency. 
Consistent results are obtained with an independent approach 
that uses minimal reactor model assumptions and 
directly predicts the far spectra from the near spectra. 
Similarly, analysis with a purely nuisance-parameter-based $\chi^2$ or
purely covariance-matrix-based $\chi^2$ yields consistent results.
The rate-only result is $\sin^{2}2\theta_{13} = \TRO $ with $\chi^2/{\rm NDF} = 0.5/4$ 
with $|\Delta m^{2}_{ee}|$ constrained
by the measurement of $|\Delta m^{2}_{\mu\mu}|$~\cite{Adamson:2013whj}.
The spectra-only result, obtained by
fixing the predicted event rate in each AD to the measured rate,
is $\sin^22\theta_{13}=\TSO$ and $|\Delta m^2_{ee}|=\DMEESO$ with $\chi^2/{\rm NDF} = 161/148$,
and rules out $\sin^22\theta_{13}=0$ at $>\!\!3$ standard deviations.

In summary, the relative deficit and spectral distortion
observed between three far and three near antineutrino detectors at Daya Bay
provides the first independent measurement of $|\Delta m^{2}_{ee}| = \DMEE $
and the most precise estimate of $\sin^22\theta_{13} = \TRS $ to date.
Following a special calibration  campaign in summer 2012,
data collection using all eight antineutrino detectors began in October 2012,
and an eventual reduction to a few percent uncertainty in both oscillation
parameters is anticipated.
On-going analysis of the special calibration data is expected to yield
improvements in the energy response model and the knowledge of the absolute \nuebar\ detection
efficiency.  These improvements will enable a future high-statistics measurement
of the absolute reactor \nuebar\ flux and energy spectra that
will provide a valuable reference for
studies of reactor neutrinos.

The Daya Bay Experiment 
is supported in part by the Ministry of Science and Technology of China,
the United States Department of Energy,
the Chinese Academy of Sciences,
the National Natural Science Foundation of China,
the Guangdong provincial government,
the Shenzhen municipal government,
the China Guangdong Nuclear Power Group,
Shanghai Laboratory for Particle Physics and Cosmology,
the Research Grants Council of the Hong Kong Special Administrative Region of China,
University Development Fund of The University of Hong Kong,
the MOE program for Research of Excellence at National Taiwan University,
National Chiao-Tung University, and NSC fund support from Taiwan,
the U.S. National Science Foundation,
the Alfred~P.~Sloan Foundation,
the Ministry of Education, Youth and Sports of the Czech Republic, 
Charles University in Prague, 
Yale University, 
the Joint Institute of Nuclear Research in Dubna, Russia,
and the NSFC-RFBR joint research program.
We acknowledge Yellow River Engineering Consulting Co., Ltd.\ and
China Railway 15th Bureau Group Co., Ltd.\ for building the underground laboratory.
We are grateful for the ongoing cooperation from the China Guangdong Nuclear
Power Group and China Light~\&~Power Company.

\bibliographystyle{apsrev4-1}
\bibliography{DYB_rateShape_prl}

\end{document}